# Machine Learning Techniques for Biomedical Image Segmentation:
# An Overview of Technical Aspects and Introduction to State-of-Art Applications

**Hyunseok Seo[1], Masoud Badiei Khuzani[1], Varun Vasudevan[2],**

**Charles Huang[3], Hongyi Ren[1], Ruoxiu Xiao[1], Xiao Jia[1], and Lei Xing[1]\***

**\*Corresponding author: lei@stanford.edu**
**Phone: +1-650-498-7896, Fax: +1-650-498-4015**

1. Medical Physics Division in the Department of Radiation Oncology, School of Medicine, Stanford University, Stanford, CA, 94305, USA

2. Institute for Computational and Mathematical Engineering, School of Engineering, Stanford University, Stanford, CA, 94305, USA

3. Department of Bioengineering, School of Engineering and Medicine, Stanford University, Stanford, CA, 94305, USA

Total words: 9313

Total Tables: 3, total Figures: 14

**Abstract**

In recent years, significant progress has been made in developing more accurate and efficient machine learning algorithms for segmentation of medical and natural images. In this review article, we highlight the imperative role of machine learning algorithms in enabling efficient and accurate segmentation in the field of medical imaging. We specifically focus on several key studies pertaining to the application of machine learning methods to biomedical image segmentation. We review classical machine learning algorithms such as Markov random fields, k-means clustering, random forest, etc. Although such classical learning models are often less accurate compared to the deep learning techniques, they are often more sample efficient and have a less complex structure. We also review different deep learning architectures, such as the artificial neural networks (ANNs), the convolutional neural networks (CNNs), and the recurrent neural networks (RNNs), and present the segmentation results attained by those learning models that were published in the past three years. We highlight the successes and limitations of each machine learning paradigm. In addition, we discuss several challenges related to the training of different machine learning models, and we present some heuristics to address those challenges.

## 1. Introduction

Segmentation is the process of clustering an image into several coherent sub-regions according to the extracted features, *e.g.*, color, or texture attributes, and classifying each sub-region into one of the pre-determined classes. Segmentation can also be viewed as a form of image compression which is a crucial step in inferring knowledge from imagery and thus has extensive applications in precision medicine for the development of computer-aided diagnosis based on radiological images with different modalities such as magnetic resonance imaging (MRI), computed tomography (CT), or colonoscopy images.

Broadly, segmentation techniques are divided into two categories (*i.e.*, supervised and unsupervised). In the unsupervised segmentation paradigm, only the structure of the image is leveraged. In particular, unsupervised segmentation techniques rely on the intensity or gradient analysis of the image via various strategies such as thresholding, graph cut, edge detection, and deformation, to delineate the boundaries of the target object in the image. Such approaches perform well when the boundaries are well-defined. Nevertheless, gradient-based segmentation techniques are prone to image noise and artifacts that result in missing or diffuse organ/tissue boundaries. Graph-based models such as Markov random fields  are another class of unsupervised segmentation techniques that are robust to noise and somewhat alleviate those issues, but often comes with a high computational cost due to employing iterative scheme to enhance the segmentation results in multiple steps.

In contrast, supervised segmentation methods incorporate prior knowledge about the image processing task through training samples [1]. Atlas-based segmentation methods are an example of supervised models that attracted much attention in the 1990s [2,3]. These types of methods, such as probabilistic atlases and statistical shape models, can capture the organs' shape well and generate more accurate results compared to unsupervised models. Support vector machine (SVM), random forest (RF), and *k*-nearest neighbor clustering are also among supervised segmentation techniques that have been studied rigorously in the past decade. However, the success of such methods in delineating fuzzy boundaries of organs in radiological images is limited.

In recent years, significant progress has been made in attaining more accurate segmentation results within the supervised framework of machine learning. In particular, deep convolutional neural networks (CNNs) have achieved the state-of-the-art performance for the semantic segmentation of natural images; see, *e.g.*, [4,5]. This success is largely due to the paradigm shift from manual to automatic feature extraction enabled by deep learning networks combined with significant improvements in computational power. Such automatic feature extraction is guided by a large amount of training data. The research trends of applying deep learning to medical image analysis was well organized by Litjens *et al.* [6], which shows that deep learning studies have been dramatically increased since 2015. The seminal paper of Litjens *et al.* [6], offers a wide range of deep learning techniques for medical image analysis. In particular, the authors summarize deep learning methods for various clinical tasks such as image classification, object detection, disease quantification, and segmentation, among many others. In contrast, the scope of this article is broader in the sense that, we review a wide range of machine learning techniques, including deep learning (*e.g.*, see [7-12]), kernel SVMs, Markov Random fields, random forests, etc. Nevertheless, we consider the applications of such machine learning techniques to medical image segmentation only, and present the evaluations results in that context.

The rest of this paper is organized as follows. In Section 2, we review classical machine learning techniques such as kernel support vector machines (SVMs), random forests, Markov random field, and present their application to the medical image segmentation. In Section 3, we present segmentation methods based on more traditional methods outside the machine learning paradigm. In section 4, we review preliminaries of the deep learning methods and present the application of different deep learning architectures to the medical image segmentation that were published in the past three years. In section 5, we discuss the limitations of current machine learning models in medical applications and we present useful strategies to circumvent those limitations.

## 2. Classical machine learning methods

### 2.1. Overview of classical machine learning

#### 2.1.1. Kernel support vector machine (SVM)

The SVMs are supervised machine learning techniques that make a non-probabilistic binary classifier by assigning new examples to one class or the other. More specifically, the kernel support vector machines (SVM) is a nonlinear classifier where the representations are built from pre-specified filters. This is in contrast to the deep learning paradigm in which good representations are learned from data.

Consequently, the kernel SVM are sample efficient learning methods that are more adequate for medical imaging applications with a small training sample size. In addition, the training phase of the kernel SVM involves tuning the hyperparameters of the SVM classifier only, which can be carried out quickly and efficiently. Contrary to deep learning models, the kernel SVM is a transparent learning model whose theoretical foundations are grounded in the extensive statistical machine learning literature; see [13] and references therein for a survey of theoretical results. Figure 1 depicts the structure of a segmentation network based on the kernel SVM. The network consists of four components:

• **Feature extraction**: Feature extraction in kernel SVM is typically carried out using a filter bank with a set of pre-specified filters. Such filter bank can generate diverse representations from input data. In addition, since the filters are not learned from data, the filter bank needs to be designed based on the underlying classification task.

• **Feature selection**: In contrast to deep learning, where features are learned and guided by training data, in kernel SVM features are quite generic and thus may not be good representations for the underlying segmentation task. In addition, there could be redundant features that increase the dimensionality of the feature vectors in the feature space and cause overfitting.

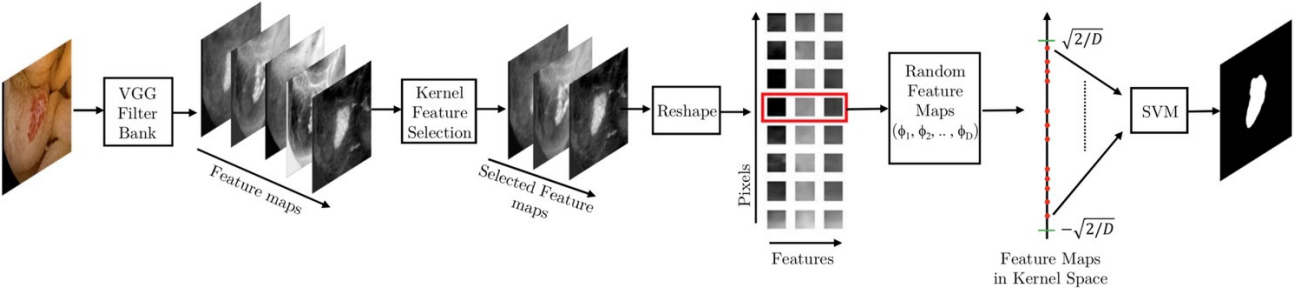

**Figure 1.** The architecture of the segmentation network based on kernel SVMs, using a filter bank in conjunction with the kernel feature selection to generate semantic representations. Random feature maps $\varphi_1, \cdots, \varphi_D$ capture the non-linear relationship between the representations and the class labels.

Feature selection algorithms are mechanisms to distill good features from redundant or noisy features. Feature selection algorithms can be supervised or unsupervised. Some examples of supervised feature selection methods are *kernel feature selection* [14], *Relief* [15], and generalized Fisher score [16]. An unsupervised feature selection using an auto-encoder is also proposed in [17].

• **Random feature maps**: At the core of kernel SVM is a kernel function that captures the non-linear relationship between the representations of input data and labels in statistical machine learning algorithms. Formally, a kernel function is defined as follows:

Let $\mathcal{X}$ be a non-empty set. Then a function $k_{\mathcal{X}} : \mathcal{X} \times \mathcal{X} \to \Re$ is called a kernel function on $\mathcal{X}$ if there exists a Hilbert space $\mathcal{H}_{\mathcal{X}}$ over $\Re$, and a map $g : \mathcal{X} \to \mathcal{H}_{\mathcal{X}}$ such that for all $x_1, x_2 \in \mathcal{X}$, we have

$$k_{\mathcal{X}}(x_1, x_2) = \langle g(x_1), g(x_2) \rangle_{\mathcal{H}_{\mathcal{X}}}, \tag{2.1}$$

where $\langle \cdot, \cdot \rangle_{\mathcal{H}_{\mathcal{X}}}$ is the inner product in the Hilbert space $\mathcal{H}_{\mathcal{X}}$.

Some examples of reproducing kernels on $\Re^d$ (in fact all these are radial) that appear throughout the paper are:

(1) Gaussian kernel: The Gaussian kernel is given by $k_{\mathcal{X}}(x, y) \equiv \exp(-\|x - y\|_2^2 / 2\sigma^2)$. .

(2) Polynomial kernel: The polynomial kernel is defined by $k_{\mathcal{X}}(x, y) \equiv (\langle x, y \rangle + c)^d$. When $c = 0$, the kernel is called homogeneous, and when $d = 1$, it is called linear.

(3) Laplacian kernel: The Laplacian kernel is similar to the Gaussian kernel, except that it is less sensitive to the bandwidth parameter. In particular, $k_{\mathcal{X}}(x, y) \equiv \exp(-\|x - y\|/\sigma)$.

The kernel methods circumvent the explicit feature mapping that is needed to learn a non-linear function or decision boundary in linear learning algorithms. Instead, the kernel methods only rely on the inner product of feature maps in the feature space, which is often known as the "kernel trick" in the machine learning literature. For large-scale classification problems, however, implicit lifting provided by the kernel trick comes with the cost of prohibitive computational and memory complexities as the kernel Gram matrix must be generated via evaluating the kernel function across all pairs of datapoints. As a result, large training sets incur large computational and storage costs.

To alleviate this issue, Rahimi and Recht proposed random Fourier features that aims to approximate low-dimensional embedding of shift invariant kernels $k_{\mathcal{X}}(x, y) = k_{\mathcal{X}}(x - y)$ via *explicit* random feature maps [18,19]. In particular, let $\varphi : \mathcal{X} \times \Xi \to \Re$ be the explicit feature map, where $\Xi$ is the support set of random features. Then, the kernel $k_{\mathcal{X}}(x - y)$ has the following

$$k_{\mathcal{X}}(x, y) = \int_{\Xi} \varphi(x, \xi) \varphi(y, \xi) \mu_{\Xi}(\mathrm{d}\xi) \qquad [2.2]$$

$$= E_{\mu_{\Xi}}[\varphi(x; \xi) \varphi(y; \xi)], \qquad [2.3]$$

where $\mu_{\Xi} \in \mathcal{P}(\Xi)$ is a probability measure, and $\mathcal{P}(\Xi)$ is the set of Borel measures with the support set $\Xi$. In the standard framework of random Fourier feature proposed by Rahimi and Rechet[18], $\varphi(x, \xi) = \sqrt{2} \cos(\langle x, \xi \rangle + b)$, where $b \sim \mathsf{Uni}[0, 2\pi]$, and $\xi \sim \mu_{\Xi}(\cdot)$. In this case, by Bochner's Theorem [20], $\mu_{\Xi}(\cdot)$ is indeed the Fourier transform of the shift invariant kernel $k_{\mathcal{X}}(x, y) = k_{\mathcal{X}}(x - y)$.

For training purposes, the expression in Eq. 3.2 is approximated using the Monte Carlo sampling method. In particular, let $\xi_1, \cdots, \xi_N \sim_{\text{i.i.d.}} \mu_{\Xi}$ be the *i.i.d.* samples. Then, the kernel function $k_{\mathcal{X}}(x, y)$ can be approximated by the sample average of the expectation in Eq. 3.3. Specifically, the following point-wise estimate has been shown in [18]:

$$k_{\mathcal{X}}(x, y) \approx \frac{1}{D} \sum_{j=1}^{D} \varphi(x; \xi_j) \varphi(y; \xi_j), \qquad [2.4]$$

where typically $D \ll n$

Using the random Fourier features $\{\varphi(x_i; \xi_j)\}_{j=1}^{n}$, the following empirical loss minimization is solved:

$$\beta^* := \arg\min_{\beta \in \Re^N} \frac{1}{n} \sum_{i=1}^{n} L\left(y_i, b + \frac{1}{\sqrt{N}} \beta^T \varphi(x_i)\right),$$
$$\text{s.t.:} \|\beta\|_{\infty} \leq R / N, \qquad [2.5]$$

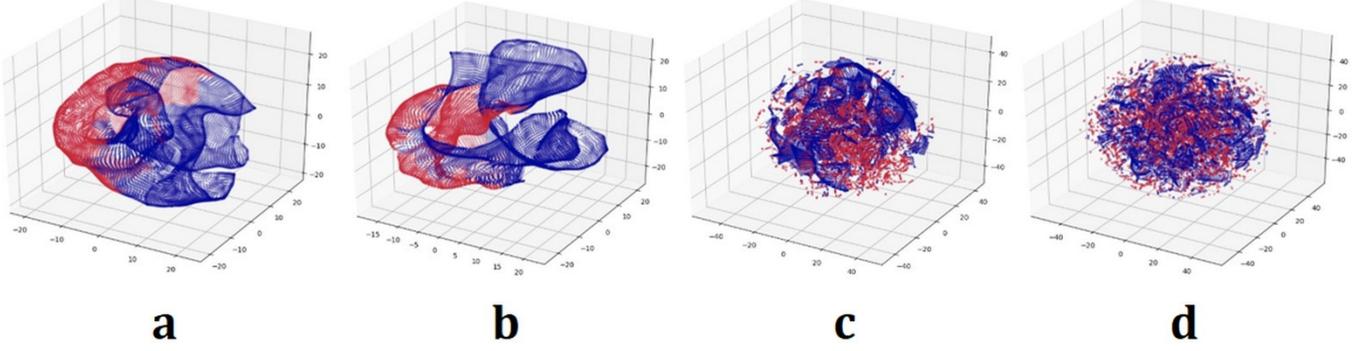

**a**      **b**      **c**      **d**

**Figure 2.** Visualization of the random feature maps in three dimensions, using the *t*-SNE plot, and for different bandwidth parameters $\gamma \equiv 1/2\sigma^2$ of the Gaussian RBF kernel $k_\chi(x,y) = \exp(-\gamma \| x-y \|_2^2)$. To generate the feature maps, the pre-trained VGG network is used. The red and blue regions correspond to the random feature maps generated by the pixels from each class label in a sampled colonoscopy image, respectively. To enhance the visualization, we have cropped the selected image and retained a balanced numbers of pixels from each class label. **(a):** $\gamma = 10^{-6}$, **(b):** $\gamma = 10^{-3}$, **(c):** $\gamma = 0.1$, **and (d):** $\gamma = 1$.

for some constant $R > 0$, where $\varphi(x) \equiv (\varphi(x, \xi_1), \cdots, \varphi(x, \xi_m))$, and $\beta \equiv (\beta_1, \cdots, \beta_D)$. Moreover $b \in \Re$ is a bias term. The approach of Rahimi and Recht [18] is appealing due to its computational tractability. In particular, preparing the feature matrix during training requires $\mathcal{O}(nD)$ computations, while evaluating a test sample needs $\mathcal{O}(D)$ computations, which significantly outperforms the complexity of traditional kernel methods.

In Fig. 2, we illustrate the three dimensional visualization of the random feature maps in the kernel space, using the *t*-SNE plot [21]. To enhance the visualization, we have cropped the selected image and retained a balanced numbers of pixels from each class label. From Fig. 2, we clearly observe the effect of the bandwidth parameter $\gamma = \dfrac{1}{2\sigma^2}$ on the accuracy of the kernel-based segmentation architecture.

In particular, as we observe from Fig. 2(c) and 2(d), choosing an unsuitable bandwidth parameters of $\gamma = 0.1$ and $\gamma = 1$ significantly degrades the classification accuracy, and results in a mixture of two classes that cannot be separated by the downstream linear SVM. The sensitivity of classification accuracy to the value of the bandwidth $\gamma$ also highlights the importance of choosing a proper bandwidth parameter for the kernel. We do not deal with such model selection issues in this review paper.

• Linear SVM: In the last layer of the segmentation network, we train a linear SVM classifier. This corresponds to the following loss function in

$$L\left(y_i, b + \frac{1}{\sqrt{N}}\beta^T \varphi(x_i)\right) = \left[1 - y_{ij}\left(b + \frac{1}{\sqrt{N}}\beta^T \varphi(x_i)\right)\right]_+ ,\qquad [2.6]$$

Where $[x]_+ = \max(0, x)$. Given a new input image $\tilde{f} = (\tilde{f}_{ij})_{(i,j)\in[I]\times[J]}$ with the feature maps $\tilde{x}_{ij}$, we generate a class label $\tilde{y}_{ij} \in \{-1, +1\}$ using

$$\tilde{y}_{ij} = \text{sgn}\left[\sum_{k=1}^{D} \beta_k^* \varphi(x_{ij}, \xi_k) + b^*\right], \qquad [2.7]$$

where $\text{sgn}$ is the sign function.

### 2.1.2. Random forest

Random forests or random decision forests are an ensemble learning method that are used to build predictive models by combining decisions from a sequence of base models. Ensemble methods use multiple learning models to gain better predictive results. In the case of a random forest, the model creates an entire forest of random uncorrelated decision trees to arrive at the best possible answer. Such methods are often called Bootstrap Aggregation or bagging, and are used to overcome a bias-variance trade-off problem. In general, learning error can be explained in terms of bias and variance. For example, if the bias is high, test results are inaccurate; and if the variance is high, the model is only suitable to certain dataset (*i.e.,* overfitting or instability). Given training dataset $X = \{x_1, \cdots, x_n\}$ with labels $Y = \{y_1, \cdots, y_n\}$, bagging repeatedly and randomly samples ($K$ times) the training dataset, and replaces the original training dataset by fitting binary trees to these samples. Let $X_k$ and $Y_k$ be the sampled dataset, where $k = \{1, \cdots, K\}$, and let $T_b$ denote the binary tree trained with respect to $X_k$ and $Y_k$. After training, predictions on the test dataset, $\tilde{x}$, can be made in two ways:

• **Averaging the predictions from all individual trees**: $\tilde{y} = \dfrac{1}{K} \sum_K T_b(\tilde{x})$

• **Taking the majority vote in the case of classification trees**

The bias in learning error reduces by averaging results from respective trees, and while the predictions of a single tree are highly sensitive to its training set, the mean of individual trees is not sensitive, as long as the trees are not correlated. If trees are independent from each other, then the central limit theorem would ensure variance reduction. Random forest uses an algorithm which selects a random subset of the features at the process of splitting each candidate to reduce the correlation of the trees in a bagging sample [22]. Another advantage of random forest is that it is easy to use, and requires tuning only three hyperparameters, namely, the number of trees, the number of features used in a tree, and the sampling rate for bagging. Moreover, the results from random forest have a high accuracy with stability, however, the internal process of it is a kind of black box like deep learning.

### 2.1.3. Linear regression

Linear regression is perhaps one of the most well-known methods in statistics and machine learning, whose theoretical performance is studied extensively. Despite its simple framework, its concept is still a basis for other advanced techniques. In linear regression, the model is determined by linear functions whose unknown parameters are estimated from data [23]. Simply put, linear regression is related to finding a linear equation which represents the model well. Linear regression models are often fitted using minimization of the $l$-norm (ex., 2-norm minimization is the least square approach).

### 2.1.4. Markov random field (MRF)

Another segmentation method using the classical machine learning concept is the Markov random field (MRF) segmentation. MRF is itself a conditional probability model, where the probability of a pixel is affected by its neighboring pixels. MRF is a stochastic process that uses the local features of the image [24,25]. It is a powerful method to connect spatial continuity due to prior contextual information. So, it provides useful information for segmentation. A brief summary of the MRF is well described by Ibragimov and Xing [26]: According to MRF formulation, the target image can be represented as a graph $G = \{V, E\}$, where $V$ is the vertex set and $E$ is the edge set. A vertex in $G$ represents a pixel in the images and an edge between two vertices indicate that the corresponding pixels are neighbors. For each object $S$ in the image, each vertex is assigned with label 1 when it belongs to $S$, and with label 0 when it does not. Then, the label of a voxel is, finally, determined by a its similarity to object $S$ (*i.e.*, probability $P_x^S$) and similarity to object $S$ of each neighbors.

### 2.2. Segmentation results of medical images from classical machine learning

The classical machine learning algorithms, such as SVM, Random forest, or MRF, were applied to classical medical image segmentation [25,27-31] with nice results. Held *et al.*[25] probably first introduced the segmentation method using Markov random field to address the following three practical issues on MR images simultaneously. Their segmentation algorithm captures three key features that are practical obstructers to MR image segmentation (*i.e.*, nonparametric distributions of tissue intensities, neighborhood correlations, and signal inhomogeneities):

• Nonparametric distribution of tissue intensities are modeled by Parzen-window [32] statistics.

• Neighbor tissue correlations are dealt with MRF to manage the noisy MR data.

• Signal inhomogeneities are also described by a priori MRF.

Then, the statistical model is optimized by simulated annealing or iterated conditional modes. They offered the segmentation of simulated MR images with respect to noise, inhomogeneity, smoothing, and optimization method. The accuracy was measured by error rate and the error rates in most cases were less than 10 %.

In Fig. 3 and Fig. 4, we illustrate the segmentation results for four sampled images from the GIANA challenge dataset, using FCN [33] and the kernel SVM with a scattering network [34] in Fig. 2. We train both networks on one percent of the dataset to showcase the ability of the kernel SVM architecture in adapting to small training sample sizes.

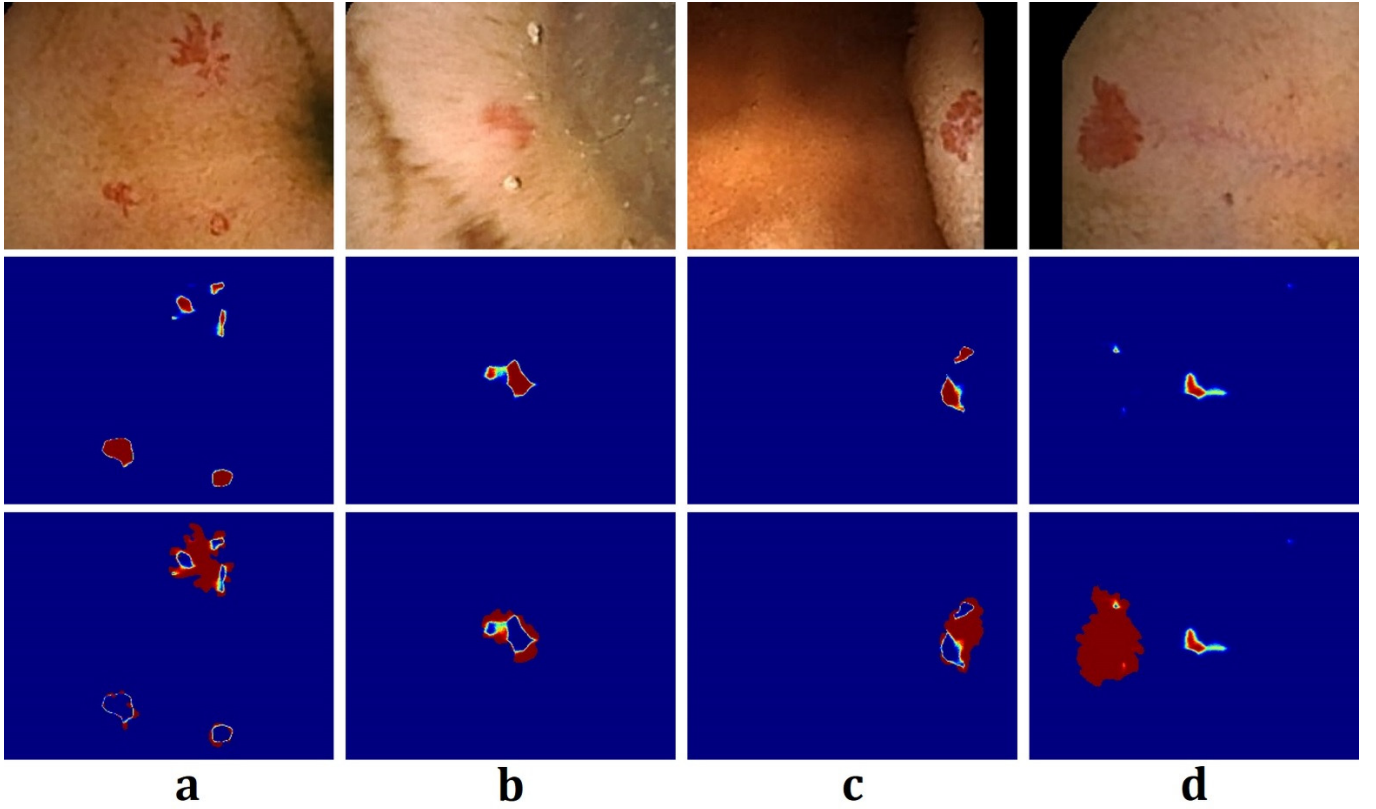

**Figure 3. Segmentation of Angiodysplasia colonoscopy images generated by FCN on sampled test images from the GIANA challenge dataset. Top: the colonoscopy images obtained using Wireless Capsule Endoscopy (WCE), Middle: the heat maps depicting the soft-max output of FCN, Bottom: the heat map of the residual image computed as the absolute difference between the proposed segmentation and the ground truth. Due to training on a small data-set, FCN tends to overfit and does not generalize well to unseen data.**

Figure 3 shows the segmentation results, using the FCN architecture. The middle row corresponds to the heat map generated from the soft-max output of the FCN. In addition, the bottom row shows the heat map of the residual image, computed as the absolute difference between the generated segmentation map and the ground truth. From Figs. 3(a-c), we observe that while FCN correctly locates the swollen blood vessels from the surrounding tissues, the segmentation results is rather poor as can be seen in the bottom row of Fig. 3. In the case of Fig. 3(d), the FCN almost entirely misses the swollen blood vessels. Figure 4 illustrates the segmentation results for the same images using the kernel SVM architecture. Here, the heat maps are generated via the soft-max function (*a.k.a.* the inverse logit function) of the kernel SVM classifier, *i.e.*, for each pixel, we generate the output

$$logit^{-1}\left(\frac{1}{\sqrt{D}}\beta^T\varphi(x)\right)\equiv\frac{\exp\left(\frac{1}{\sqrt{D}}\beta^T\varphi(x)\right)}{1+\exp\left(\frac{1}{\sqrt{D}}\beta^T\varphi(x)\right)}.$$  [2.8]

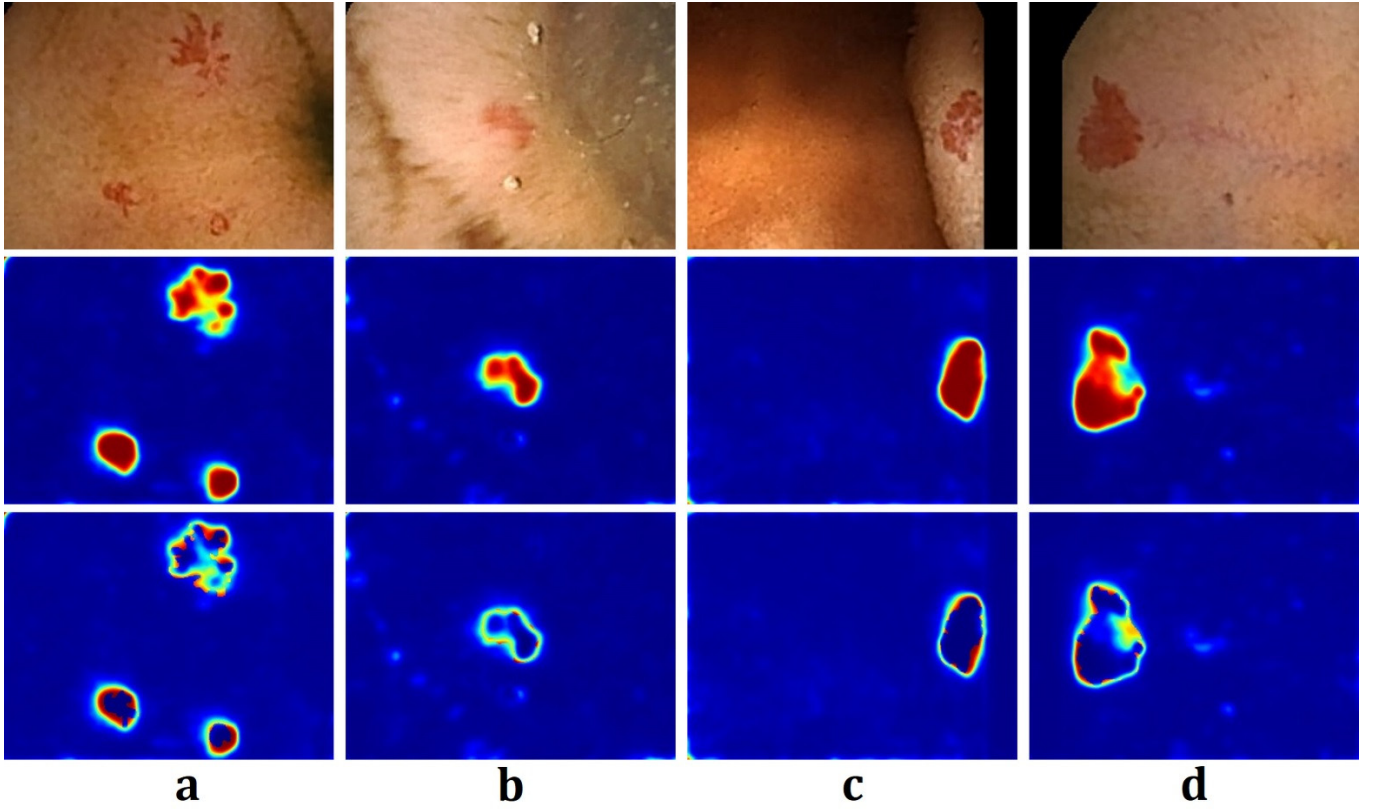

**Figure 4.** Segmentation of Angiodysplasia colonoscopy images on sampled test images from the GIANA challenge dataset, generated via the kernel SVM using the VGG filter bank with the kernel feature selection. The bandwidth of RBF kernel $1/2\varrho^2$ is selected via maximum mean discrepancy optimization. **Top:** the colonscopy images obtained using Wireless Capsule Endoscopy (WCE), **Middle:** the heat maps depicting the soft-max of SVM kernel classifier, **Bottom:** the heat map of the residual image computed as the absolute difference between the proposed segmentation and the ground truth. Despite training on a small data-set, the kernel SVM performs well on the test data set.

We observe from Figs. 3 and 4 that the segmentation results from the kernel SVM outperforms those of FCN. Moreover, while FCN misses the bleeding region in Fig. 3(d), the SVM network generates correct segmentation.

In Fig. 5, we illustrate the jitter plots as well as box plots for the mean IoU scores defined as

$$M_{\mathrm{IoU}} \equiv \frac{1}{2}\frac{n_{11}}{n_{12}+n_{21}+n_{11}} + \frac{1}{2}\frac{n_{22}}{n_{12}+n_{21}+n_{22}}, \qquad [2.9]$$

where $n_{ij}$ be the number of pixels of class $i$ predicted to belong to class $j$. We compute $M_{\mathrm{IoU}}$ for both the kernel SVM network as well as FCN on the test dataset.

We use different numbers of training samples to evaluate the performance of each architecture, as demonstrated in Fig. 5. We observe that on a small training dataset, the kernel SVM achieves higher IoU scores than the deep learning network. This is due to the fact that fewer hyperparameters are need to be determined during the training phase of the kernel SVM. In contrast, due to the large number of hyperparameters that must be determined in FCN from a small training sample size, the network is prone to overfitting, even with regularization techniques such as dropout.

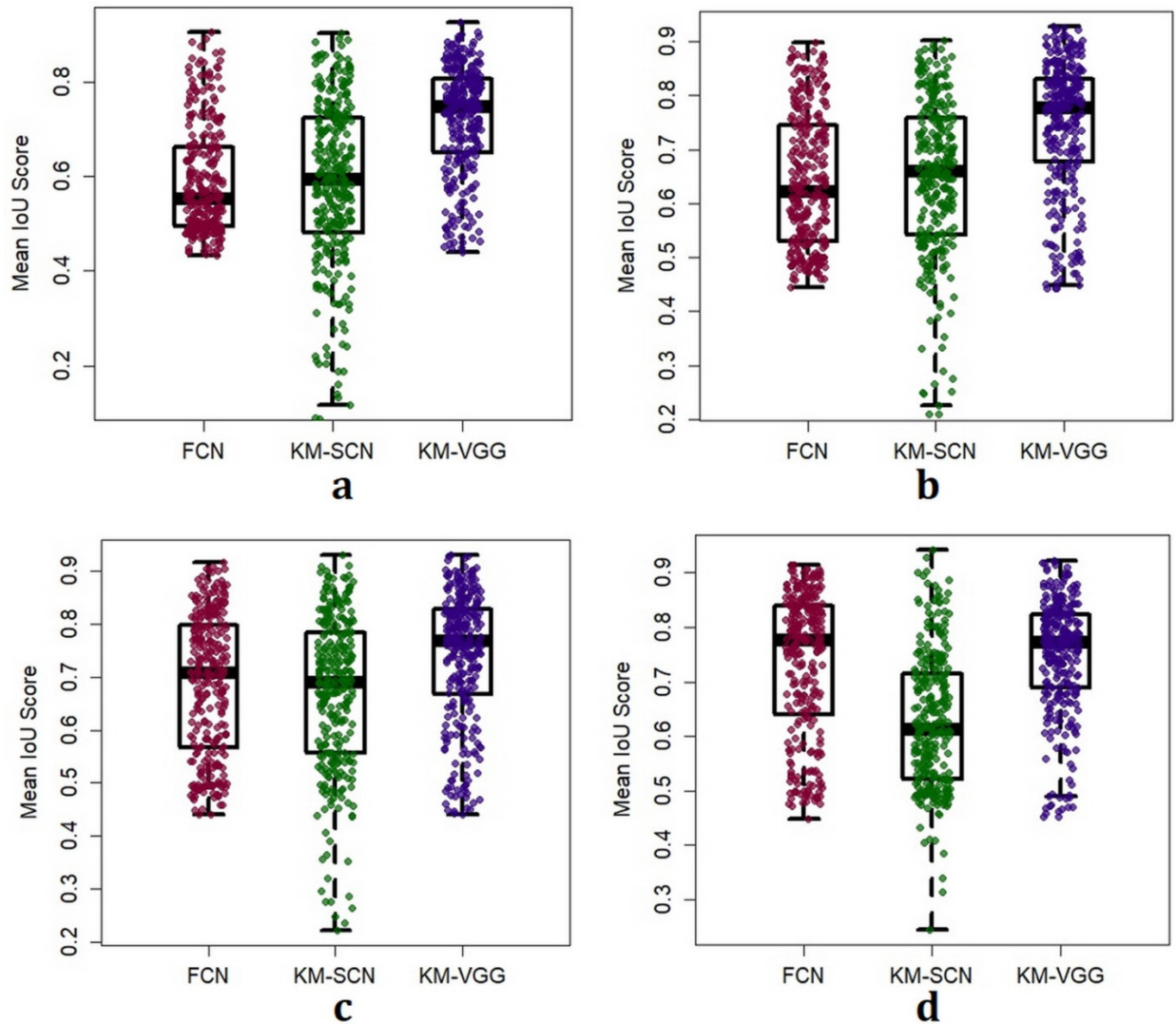

**Figure 5.** Comparison of the mean IoU score $M_{\text{IoU}}$ for FCN (the red color), the kernel SVM with Mallat's scattering network as the filter bank (the green color), and the kernel SVM with a pre-trained VGG network as a filter bank (the blue color) on the test dataset. To tune the parameters of the kernel in the Gaussian RBF kernel, the two-sample test is performed. Each plot correspond to the performance of networks that are trained on different sample sizes. Panel (a): 76800 Pixels (1 image), Panel (b): 153600 Pixels (two images), Panel (c): Trained on 1 % of the data-set (3 images), (d): Trained on 5 % of the data-set (15 images).

From Fig. 5, we also observe that increasing the training sample size does not change the performance of the kernel SVM significantly as the hyperparameters of the classifiers converge to their optimal values very quickly with a few training samples. In contrast, due to the large representational capacity of deep learning network and due to a large number of hyperparameters in the network, increasing the number of training samples significantly improves the performance of FCN.

## 3. Other related segmentation methods

## 3.1. Overview of other related segmentation methods

### 3.1.1 Atlas-based segmentation

Atlas-based segmentation, strictly speaking, does not belong to general machine learning algorithms, but is a specific method for segmentation with high performance. Rohlfing *et al.* [35] mathematically described atlas-based segmentation in detail: An atlas $A$ is a mapping $A : \Re^n \to L$ from $n$-dimensional spatial coordinates to labels. Conceptually, an atlas is similar to mapping from $\Re^n$ to the space of gray values that is subset of $\Re$, so atlas can itself be considered as a special type of image, *i.e.*, a label image. To apply an atlas $A$ to a new image, $S$, registration should be performed for coordinate mapping. An atlas is usually generated by manual segmentation, which can be expressed as a mapping, $M : \Re^n \to \Re$. For image segmentation of $S$ based on atlas, each point in an image has a corresponding equivalent in the other. This correspondence of two images can be represented as a coordinate transform $\mathbf{T}$ that maps the image coordinates of $S$ onto those of $M$. Then, for a given position $x$ in $S$, we can find a corresponding label $x$ as follows,

$$x \to A(\mathbf{T}(x)) . \qquad\qquad [3.1]$$

The transformation of $\mathbf{T}$ is determined by image registration.

### 3.1.2. Deformable model segmentation

Deformable model segmentation is also a specific method for segmentation. Deformable models are implemented as curves or surfaces like physical bodies that have certain elasticity trying to keep their shape, while the image that we want to segment is represented as potential field with force that deforms the model to delineate object shape, minimizing a cost function [36,37]. The force is defined with an internal and external force. Internal force works to preserve the shape smoothness of the model, whereas, the external force is related to image features for desired image boundaries. The representative deformable model segmentation is widely known as an active contour whose deformations are determined by the displacement of a finite number of control points along the contour [37].

### 3.1.3. Superpixel-based segmentation

Superpixels are perceptually meaningful image regions generated by grouping pixels. They are commonly used in segmentation algorithms as a preprocessing step. Once superpixels are formed, they are used as the basic processing units for the subsequent segmentation task. A good superpixel algorithm should improve the performance (both speed and quality of the results) of the segmentation algorithm that uses it [38]. Algorithms for generating superpixels can be categorized into graph-based, gradient-ascent based, K-means clustering-based and entropy-rate-based methods [39,40]. Tian *et al.* [41] proposed a superpixel based 3D graph cut algorithm to segment the prostate on magnetic resonance images. The superpixels are usually combined with other machine learning techniques as well [42,43].

3.2. Segmentation results of medical images from other related methods

Prior to modern advances in deep learning methods, atlas-based and deformable model segmentations were one of the most popular methods for medical images, and their results were well described by Xu et al. [44] and Cabezas et al. [45]. Nikolov et al.[46] organized current performance of atlas- and deep learning-based segmentation, which shows some atlas-based segmentation methods have more accurate segmentation results than those from deep learning-based methods (98.0 % vs. 94.0 % for mandible). Ji et al. [42] applied superpixels to the segmentation of MR brain image, and Tian et al. [41] proposed a superpixel-based 3D graph cut algorithm for segmenting the prostate on MR images. The superpixels instead of pixels were considered as a basic unis for 3D graph cut, and they also used a 3D active contour model to overcome the drawback of graph cut, like smoothing. By doing this, they achieved the a mean DSC of 89.3 %, which was the highest score. Irving et al. [43] introduced a simple linear iterative clustering for superpixels within region of interest and showed better representation of brain-tumor sub-regions. Now they have been combined with deep learning [26,47,48].

**4. Deep learning methods**

Before starting a review of the deep learning, we summarize the key terminologies used throughout this section in Table 1.

| Terminology | Description in the manuscript |
|---|---|
| Receptive field | Region that can possibly influence the activation |
| Selective window | Selective pixel region |
| Overfitting | Result is too sensitive to certain datasets |
| Feed-forward network | Network in which the input data goes through many hidden layers and finally reaches the output layer |
| Hyperparameter | Parameters whose values are not set before learning process |
| Stride | Amount by which the convolution kernel shifts |
| Atrous | Distance between kernel elements (weights) |
| Pooling | Reduction of signal dimensionality in the individual network layers |
| Activation function | Point-wise non linearity with respect to input value |
| Back propagation | Propagation of the loss back into the network to update weights via gradient descent approach that exploits the chain rule. |

**Table 1. Definitions of deep learning terminologies.**

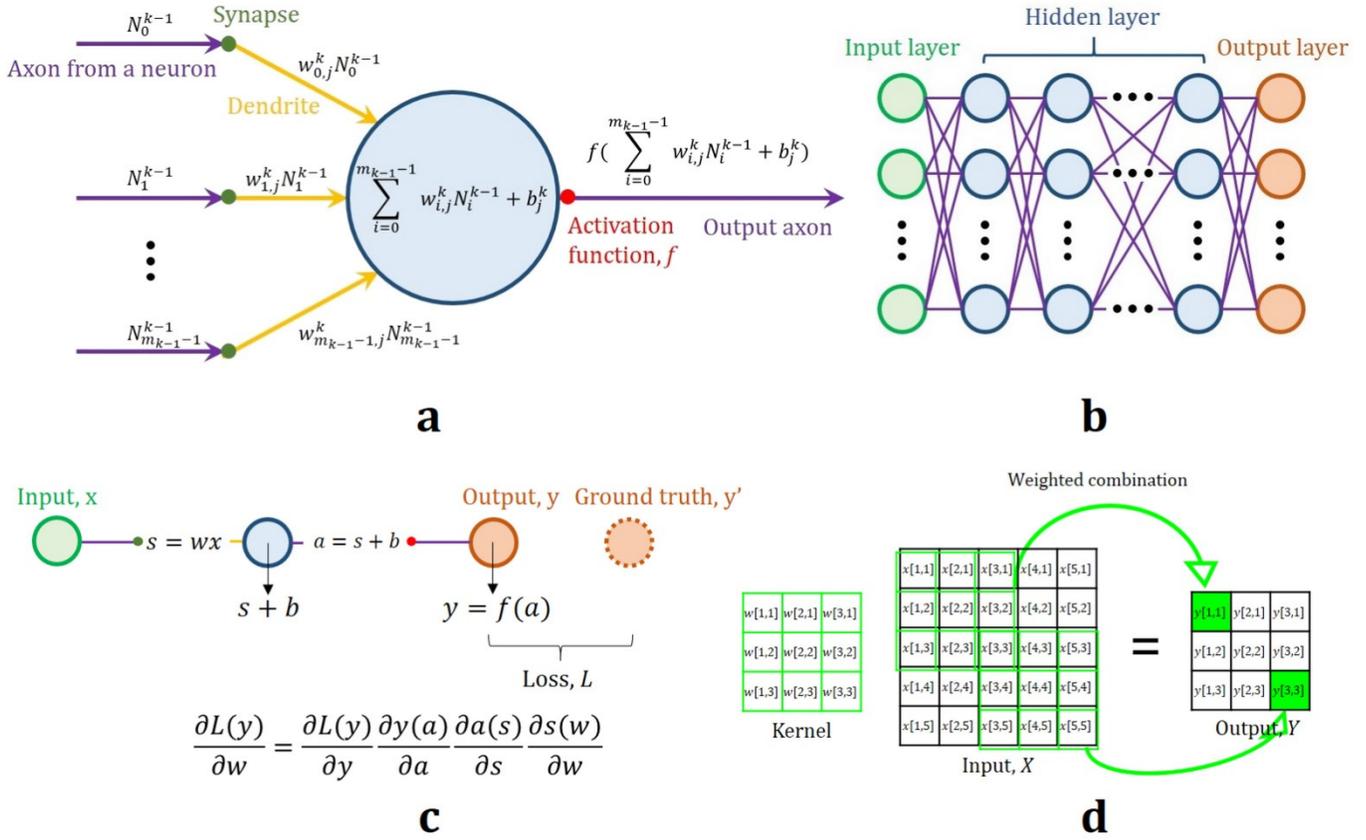

**Figure 6. The architecture of the artificial neural network (ANN). (a) Mathematical model of a perceptron (node). (b) Multi-layer perceptron (MLP) structure for ANN. Each node in the hidden layer of (b) is described mathematically in (a). (c) An example of back-propagation. Loss is minimized by the update of the weight, $w$ based on the gradient of the loss function with respect to $w$ via the chain rule where $b$ is the constant bias. (d) An example of convolution operation in CNN. Same kernel weights are applied to convolution operation for an output.**

## 4.1. Overview of deep learning networks

### 4.1.1. Artificial Neural Network (ANN)

Basic network model of deep learning is the ANN, which is fully connected from input to output by cascading perceptrons, as shown in Fig. 6. The first concept of artificial neurons was described by McCulloch and Pitts [49], which was developed into perceptron posited [50] in 1958. The node (perceptron) in Fig. 6(a) has a mathematical model that can express signal transfer similar to the biological neuron. Output of the $j^{\text{th}}$ node in the $k^{\text{th}}$ layer, $N_j^k$, is defined as follows,

$$N_j^k = f(\sum_{i=0}^{m_{k-1}-1} w_{i,j}^k N_i^{k-1} + b_j^k), \qquad [4.1]$$

where $w_{i,j}^k$ is the weighting value of the $i^{\text{th}}$ output of node in the $k-1^{\text{th}}$ layer for the $j^{\text{th}}$ node in the $k^{\text{th}}$ layer, $b_j^k$ is a constant bias value for the $j^{\text{th}}$ node in the $k^{\text{th}}$ layer, $f(\cdot)$ is the activation function of $\cdot$ for imposing non linearity to the network, and total number of nodes in the $k-1^{\text{th}}$ layer is $m_{k-1}$. The network is composed of multiple nodes connected to each other, as shown in Fig. 6(b). The weights and bias values are updated via back-propagation principle during training to reduce the predefined loss function [51-54]. Back-propagation is a way to propagate the loss between the prediction and ground truth back into the network in order to calculate the amount of update for weights. This is performed by following a gradient descent approach that exploits the chain rule from calculus. Figure 6(c) shows the simplest case of the back-propagation calculating the gradient of the loss function with respect to the weight via the chain rule. Increasing the number of hidden layers in ANN increases the flexibility of the model [55-57]. In the early 1990s, Blanz and Gish [58] showed that multi-layer perceptron (MLP) based on ANN could handle image segmentation problem. ANN based networks consider all combinations of features in previous layers, however, they are computationally expensive because of their fully connected structure [59].

### 4.1.2. Convolutional Neural Network (CNN)

Recent architectures for image segmentation most commonly use CNN to assign class labels to patches of the image. CNN was first introduced by Lecun *et al.* [51,60] and has become a dominant network architecture in computer vision and image analysis. Convolutional layers can effectively capture local and global features in images, and by nesting many such layers in a hierarchical manner, CNNs attempt to extract broader structure. Further, they allow for more efficient learning through parameter sharing, as show in Fig. 6(d). From successive convolutional layers that capture increasingly complex features in the image, a CNN can encode an image as a compact representation of its contents.

The basic building blocks of CNN consists of a convolutional transformation with a set of filters that are learned from data as well as non-linearity, and pooling operations. In what follows, we review each building block:

• Convolutional transform: The network we consider consist of $d$ convolutional layers. Each layer applies a convolutional transform that is made up of a set of unstructured filters (kernels) $\Psi_{\Lambda_n} = \{g_{\lambda_n}\}_{\lambda_n \in \Lambda_n}, n = 1, 2, \cdots, d$ to generate different representations of input image. The finite index set $\Lambda_n$ is the collection of filters in the $n^{th}$ layer.

• Pooling operation: The pooling operation reduces signal dimensionality in the individual network layers and ensures robustness of the feature vector with respect to deformations and translations. A Lipschitz-continuous pooling operator $P_n : \mathfrak{R}^{N_n} \times \mathfrak{R}^{N_n} \to \mathfrak{R}^{N_n/S_n} \times \mathfrak{R}^{N_n/S_n}\$,$ where the integers $S_n \in \mathbf{N}$, with $N_n / S_n \equiv N_{n+1} \in \mathbf{N}$ is referred to as pooling factor. Some examples of pooling operations are as follows,

(1) Sub-sampling: This operation amounts to $P_n : \mathfrak{R}^{N_n} \to \mathfrak{R}^{N_{n+1}}, (Pf)[m] = f[S_n m]$. When $S_n = 1$, $(Pf) = f$ is the identity operator.

(2) Average pooling: This is defined as $P_n : \mathfrak{R}^{N_n} \to \mathfrak{R}^{N_{n+1}}, (Pf)[m] = \frac{1}{S_n} \sum_{k=S_n m}^{S_n m + S_n - 1} f[k]$ for $m \in \{0, 1, 2, \cdots, N_{n+1}\}$.

(3) Max pooling: This is defined by $P_n : \Re^{N_n} \to \Re^{N_{n+1}}$,

$(Pf)[m] = f[S_n m] = \max_{k \in \{S_n m, \cdots, S_n m + S_n - 1\}} |f[k]|$, for $n \in \{0, 1, 2, \cdots, N_{d+1}\}$.

• Non-linearity (or activation): A point-wise non-linearity $\rho_n : \Re \to \Re$ that is Lipschitz $|\rho(x) - \rho(y)| \leq L |x - y|, \forall x, y \in \Re$ is applied after each convolution layer. Some examples of non-linearities are as follows:

(1) Hyperbolic tangent: The non-linearity is defined as $\rho(x) = \dfrac{e^x - e^{-x}}{e^x + e^{-x}}$, and has the Lipschitz constant $L = 2$.

(2) Rectified linear unit (ReLU): The non-linearity is defined by $\rho(x) = \max\{0, x\}$ with Lipschitz constant $L = 1$.

(3) Modulus: The non-linearity is defined as $\rho(x) = |x|$, and has the Lipchitz constant $L = 1$.

We remark that ReLU non-linearity was initially introduced by Nair and Hinton [61] to circumvent gradient-vanishing problems in back-propagation algorithm. Some modifications of ReLU like leaky ReLU [62] and parametric ReLU [63] are shown to improve the classification accuracy of CNN. Weight sharing and translational invariance of CNNs significantly reduce the number of learning parameters and decrease the computation complexity. In a CNN, pooling is introduced to increase the receptive field, which is the region that can possibly influence the activation, by reducing the size of the image. The max pooling operation, which adapts the maximum value within the selective window (*i.e.*, selective pixel regions) and helps to extract more robust features, is commonly applied. At the end of the CNN, similar to ANN, a fully connected layer usually follows, which takes the weighted sum of the outputs of all previous layers to combine features that could represent the final desired output. During the network training, the weights and bias values are updated by back-propagation to minimize the predefined loss function as in the ANN [51-54].

Segmentation methods based on deep learning can be handled by supervised learning with adequate training data [64 65 66]. To build a reliable segmentation model, a prerequisite is the availability of a large amount of labeled training data. In practice, medical data is generally scarce and curation of annotated data has been one of the bottleneck problems in the widespread use of supervised deep learning in medicine.

To put the matter into perspective, the Kaggle 2017 Data Science Bowl to detect tumors in CT lung scans consists of a dataset of approximate2000 patient scans [67] whereas ImageNet large scale visual recognition challenge (ILSVRC) 2017 is composed of over 1 million natural images across 1000 object classes [68]. An important strategy to alleviate the problem is through transfer learning, which is used in deep learning to transfer the weights of a network trained on a different but related dataset. When large training data is scarce, transfer learning is a viable option for task specific model training. Generally, transfer learning proceeds either with a pre-trained model as a feature extractor for the task under study, or even more dramatically, by fine-tuning the weights of the pre-trained network while replacing and retraining the classifier on the new dataset. In the former case of transferred learning, one removes the last full connected layer, and treats the other layers as a fixed feature extractor to adapt to a new task. This strategy only trains a new classifier instead of the entire network, significantly speeding up the training process.

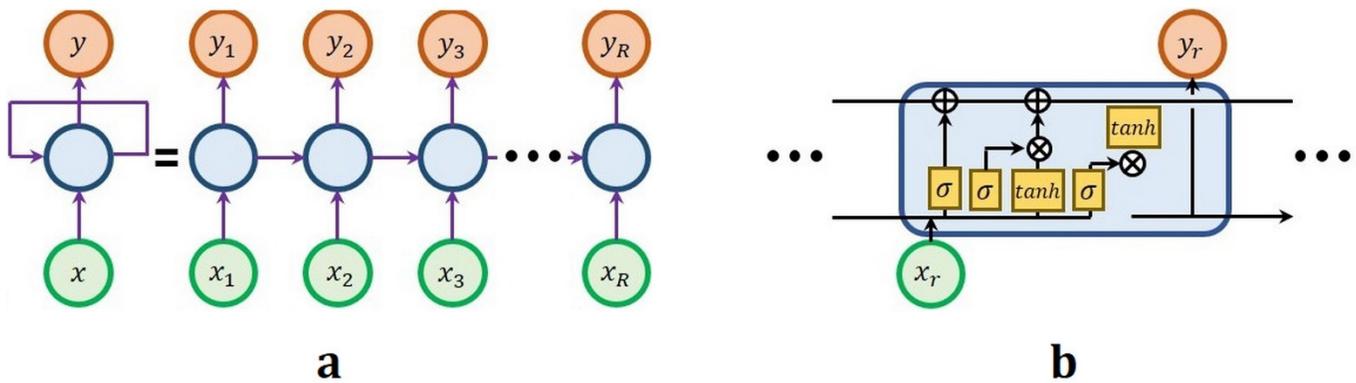

**Figure 7. The architecture of the recurrent neural network (RNN).**

Transfer learning in medical image analysis is an active area of research, especially in the past few years. Yuan *et al.* [69] developed an effective multi-parametric MRI transfer learning for autonomous prostate cancer grading. Ibragimov *et al.* [70] applied transfer learning to enhance the predictive power of a deep learning model in toxicity prediction of liver radiation therapy. The use of transfer learning for segmentation using deep learning was reported by Tajbakhsh *et al.* [71]. They applied transfer learning to segment layers of the walls in the carotid artery on ultrasound scans with pre-trained weights from Ravishankar *et al.* [72]. It was also noted that the performance of CNN can be improved by using more layers in the neural network, and the optimal number of layers may be application specific. Ghafoorian *et al.* [73] introduced the transfer learning methodology to domain adaptation of models trained on legacy MRI data that contained brain white matter hyper-intensities.

### 4.1.3. Recurrent neural network (RNN)

CNN is a feed-forward network, in which the input data goes through many hidden layers and finally reaches the output layer. Whereas, RNN is a special network where the input can be affected by the output through a recurrent path, as shown in Fig. 7(a). The feedback from output into new input can be in a role of memory that serves the connectivity of sequential data. The success of RNN depends on previous information avoiding the gradient-vanishing problem.  Long short-term memory (LSTM) for RNN was introduced [74] to effectively memorize previous information in the network. LSTM is a series of a cell states, as shown in Fig. 7(b), and the cell state has three roles to determine how much previous information is reflected in the current cell at the *forget gate*, how much current information is allowed based on previous information in the current cell at the *input gate*, and how much the output of current cell based on previous and current information is sent to next cell state at the *output gate*. The gated recurrent unit (GRU), which is modified type of LSTM, is also a popular variation of the RNN [75]. RNN is mainly used in segmentation tasks for the medical image analysis, because, if we assume that the pixel arrays along the spatial direction as the sequential input to the RNN, then the recurrent path helps to classify the current pixel based on the results of classifying previous pixels. In other words, sequential object-connectivity (morphology) information is used more relative to in CNNs.

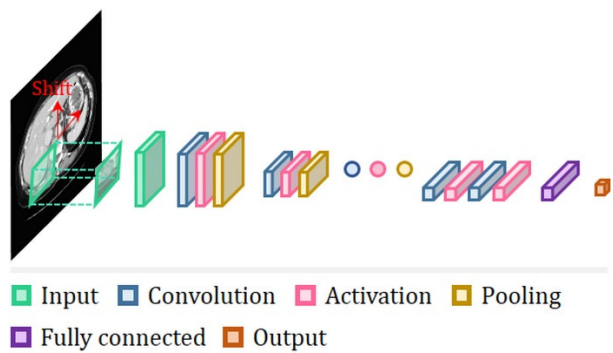

**Figure 8. Network architecture of the patch-wise CNN for liver/liver-tumor segmentation**

## 4.2. Segmentation results of medical images from deep learning

Partitioning a digital image into multiple segments for various applications has been a basic task in computer vision and medical image analysis. Numerous research and review articles have been devoted to the topic over the years. Similar to the reference [76], here we proceed by dividing previous studies on the topic into four categories:

### 4.2.1. Patch-wised convolutional neural network

Patch-based architecture is perhaps the simplest approach to train a network for segmentation. Small patches around each pixel are selected from the input images, and the network is trained by patch unit with class label pair. A schematic diagram of path-based architecture is illustrated in Fig. 8. Some popular network architectures for segmentation were designed using this approach [77-80]. The patch is usually shifted by one pixel to cover the whole image region represented in the reference [81]. Thus, it takes a long time to train the network due to the duplicated computation of pixels among neighbor patches. Another trade-off one must make is the choice of patch size and the field of view. Passing patches through numerous pooling layers results in a higher effective field of view but leads to loss of high-frequency spatial information. On the other hand, starting with small patches and using fewer pooling layers means there is less information present from which the networks can extract from. So, the patch size should be carefully chosen with consideration of specific applications. More sophisticated techniques can be applied to the input of the patch-wise deep learning networks to improve the performance on segmentation tasks.

Ibragimov and Xing [26] devised a patch-based CNN to accurately segment organs at risk (OARs) for head and neck (HaN) cancer treatment of radiation therapy. It was the first paper to demonstrate the effectiveness of deep learning for HaN cancer treatment. In particular, to achieve a good performance, the authors applied Markov random fields (MRF) as a post-processing step to merge voxel connectivity information and the morphology of OARs. The performance was evaluated on 3D CT images of 50 patients scheduled for head and neck radiotherapy, and they showed the improvement with DSCs with respect to various organs. Following the success of Ibragimov and Xing [26] in employing deep learning methods, the Google DeepMind group studied the HaN image segmentation in more detail [46]. They applied CT dataset to the 3D U-Net and achieved performance similar to experts in delineating. Qin *et al.* [47] added an object boundary class to conventional binary

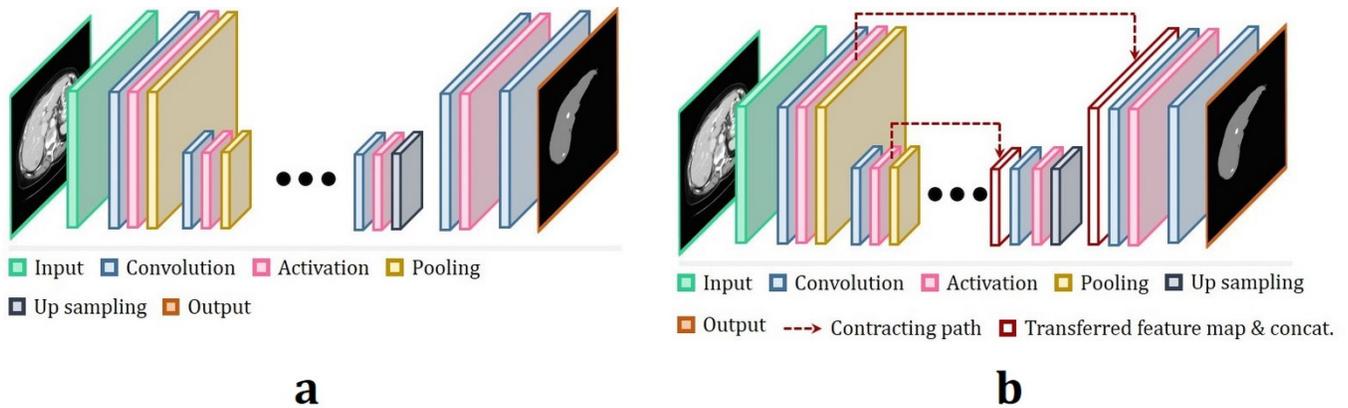

<div style="text-align:center">

**a**  **b**

</div>

**Figure 9. Network architecture of (a) FCN and (b) U-Net.**

segmentation task for object and non-object regions by preprocessing based on superpixel calculations and entropy maps. From the preprocessing of training data, three class superpixels are estimated. Then, patches are trained with three matching labels of boundary, object, and background by a patch-wise CNN. Moeskops *et al.* [82] used multiple patch sizes in the network to overcome the limitation of heuristic selection of patch size. Training is individually performed by separate networks which have different patch sizes. Only the output layer (soft-max) for the classification is shared. By doing this, hyperparameters, are optimally tuned for each patch size and corresponding kernel size.

The concept of patch-wise feature extraction can be applied to a variety of network architectures as described below.

### 4.2.2. Fully convolutional network (FCN)

FCN is different type of network architecture from the patch-wise CNN [33]. It is composed of locally connected layers such as convolution, pooling, and up pooling (up sampling). This type of network directly outputs a full-size segmentation map. It can reduce the number of hyperparameters and computational complexity due to down sampled feature maps (pooling). The basic architecture is similar to autoencoders, as shown in Fig. 9(a). The encoder part extracts the features with pooling, and the original input size recovers in the decoder part while deconvolving higher level features extracted from the encoder part. There are many studies using FCN for segmentation [83-86]. The most popular one being the U-Net [87], which consists of a conventional FCN combined with skip connections between the encoder part and decoder part, as shown in Fig. 9(b). High resolution features from the encoder part are transferred to and is combined with up sampled outputs in the decoder part by skip connections. Then, the successive convolution layer can learn more precise results by assembling the encoder and decoder parts. The original U-Net has shown superior performance for medical image segmentation tasks.

Most early deep learning approaches are only able to apply to 2D images, however, in most clinical cases, medical images are composed of 3D volumetric data. Similar to the U-Net, the V-Net is a new architecture for 3D segmentation based on 3D CNN [88]. The V-Net uses 3D convolutions to ensure the correlation between adjacent slices for feature extraction. The V-Net has another path connecting the input and the output of each stage to enable learning of residual values [89]. In general, 3D volumetric data size requires a large amount of memory. The author of the V-Net paper also noted that, depending on the specific implementation, replacing pooling operations with convolution operations can save system memory, because mapping the output of pooling back to input is not needed anymore in the back-propagation step. In addition, replacing pooling operations can be better understood and analyzed [90] by applying only deconvolutions instead of up pooling operations. A number of papers using U-Net and V-Net architectures for segmentation have been published [91-93 94]. It is perhaps worth of noting that, according to Salehi *et al.* [95], FCN may cause data imbalance due to the use of entire samples to extract local and global image features. For example, in the case of lesion detection, the number of normal voxels is typically 500 times larger than that of lesion voxels. *Salehi et al.* [95] proposed new loss function based on Tversky index to reduce the imbalance through handling much better trade-off between precision and recall.

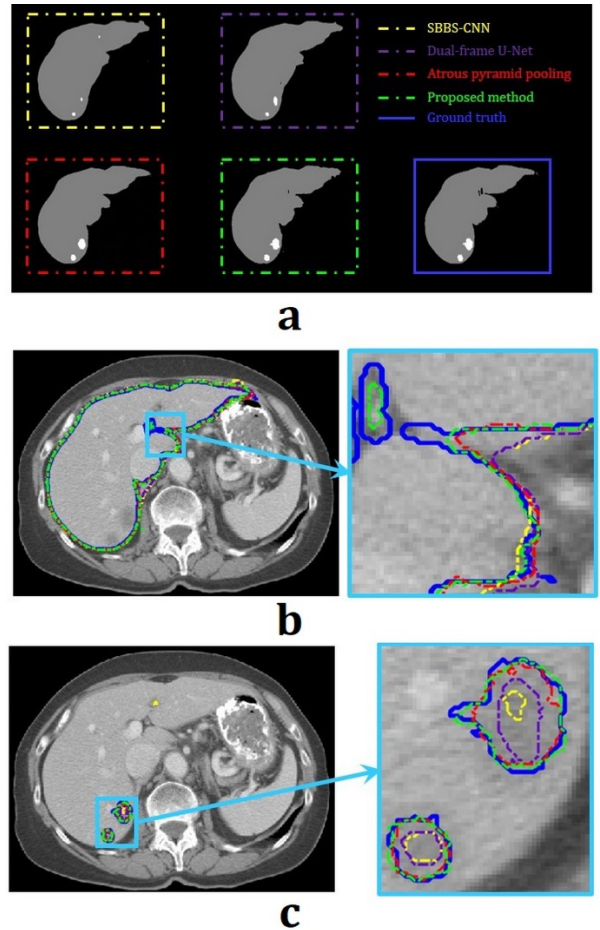

**Figure 10. (a) The results of the liver and liver-tumor segmentation. Yellow, purple, red, green, and blue lines are acquired from SBBS-CNN, dual-frame U-Net, atrous pyramid pooling, the proposed network, and ground truth, respectively. (b) and (c) are the contouring of the segmentation results in (a).**

The segmentation results usually depend on the boundary information of the object. We have recently modified the conventional U-Net, which can be more sensitive to boundary information. Our network prevents duplication of low frequency component of features and extracts object-dependent high-level features. The results obtained using the modified U-Net are shown in Fig. 10. For liver-tumor segmentation, DSC of 86.68 %, volume of error (VOE) of 24.93 %, and relative volume difference (RVD) of -0.53 % were obtained. For liver segmentation, DSC of 98.77 %, VOE of 3.10 %, and RVD of 0.27 % were calculated as well. These quantitative scores are higher than the top score in the LiTS competition as of today (https://competitions.codalab.org/competitions/17094results).

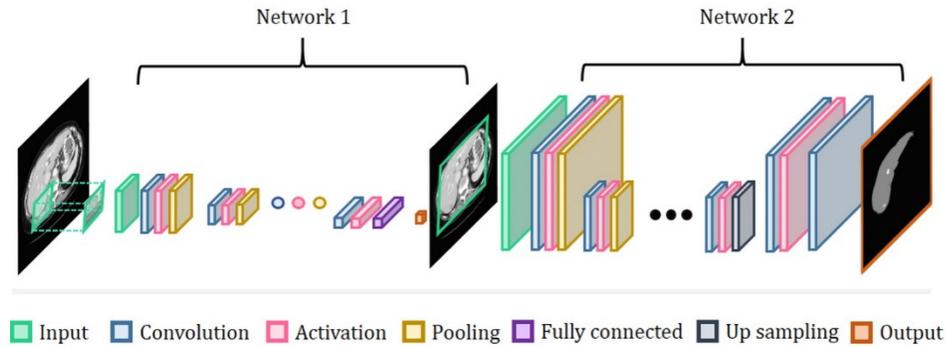

**Figure 11. Network architecture of cascaded CNN network (example of patch-wise CNN and FCN) for tumor segmentation. The first network is trained for ROI or rough classification and the second network is further tuned for final segmentation.**

### 4.2.3. Cascade multiple networks

In practice, networks are often cascaded or ensembles together. Many clinical studies involve the detection of abnormal regions and a subsequent segmentation of those regions. In these cases, cascaded networks are often used so that each subtask (*i.e.*, detection and segmentation) can be handled by a separate network and then later combined to fulfill the overall objective of the study. For instance, the first network usually focuses on detection of a region of interest (ROI), and the second performs a pixel-wise classification of the ROI into two classes (in the case of binary segmentation) or multiple classes (in the case of multi-class segmentation). In other words, rough classification is performed in the first network and results of the first network are further tuned by the second network [96,97], as shown in Fig. 11. Most of medical images are represented by gray level (one channel) unlike natural images with RGB colors (three channels). Sometimes, it causes lack of information due to low dimensionality intensity. Thus, this type of network can be powerful when there are similar structures or intensity levels in surrounding tissues. Recent works such as AdaNet build on the idea of ensemble networks and attempt to automatically select and optimize the ensemble subnetworks [98].

### 4.2.4. Other methods

The concept or shape of the fourth categorized network architecture is different from previous three network architectures. Chen *et al.* [96] combined CNN and RNN to segment the neuronal and fungal structures from 3D electron microscope (EM) images. Stollenga *et al.* [99] proposed RNN for the segmentation of 3D MRI brain images and 3D EM neuron images. Most current segmentation methods from RNN are based on the LSTM concept. Yang et al. [100] tried to apply RNN to prostate segmentations.

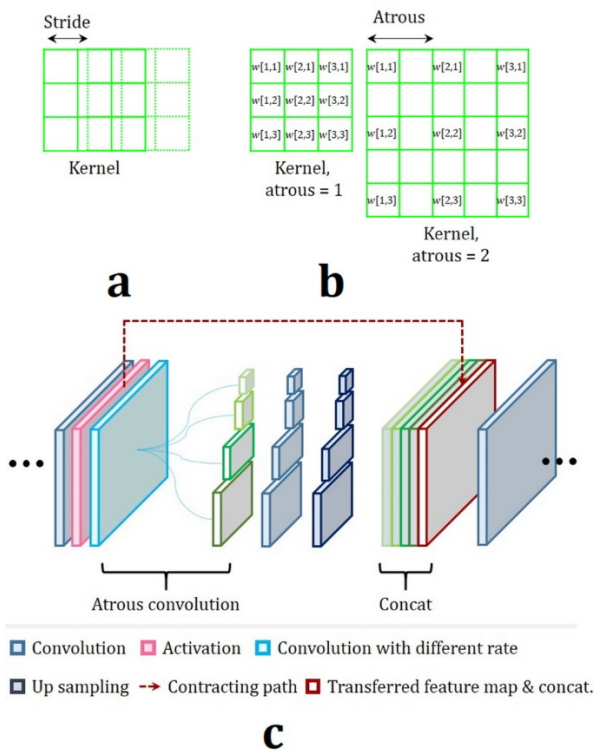

Figure 12. Descriptions of (a) stride and (b) atrous. Stride is the amount by which the convolution kernel shifts, and atrous is the distance of kernel elements (weights). (c) Structure of atrous pyramid pooling. Pyramid pooling can form the feature map which contains both local and global context information by applying different sub-region representations followed by up sampling and concatenation layers.

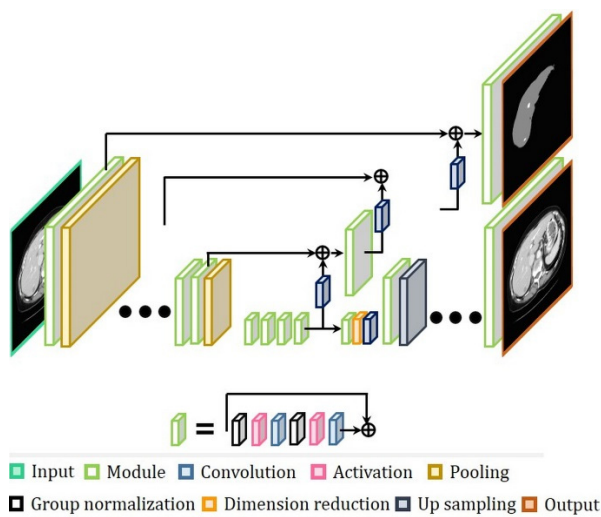

Figure 13. The network architecture ranked 1st in BRATS challenge in 2018.

Especially, for segmentation of dynamic imaging, combination of CNN and RNN can be good solution due to joint modeling of spatial and temporal information [101]. One thing to keep in mind, when RNN is used in medical image segmentation, is to apply regularization to the network. Most medical image dataset is not enough to build the deep network, so it is easy to occur the overfitting problem, which means that the result is too sensitive to certain datasets. To avoid the overfitting problem, regularization such as weight decay [102], dropout [103], and batch normalization [104] is commonly used in feed-forward network. However, conventional regularization algorithms for feed-forward network cause performance degradation of RNN, and Zaremba *et al.* [105] introduced the regularization for RNN where dropout was applied to only non-recurrent connection. By doing this, regularization can be performed without loss of previous important information. Chen *et al.* [106] proposed DeepLab architecture which is composed of up-sampled filter, atrous spatial pyramid pooling, and fully-connected Conditional Random Fields (CRF). Spatial pyramid pooling of DeepLab architecture, as shown in Fig. 12, prevents information (resolution) loss from the conventional pooling used to enlarge receptive field, so it has been applied to medical image processing to segment lesion by localizing object boundary clearly [107,108]. Myronenko [109] developed a deep learning network 3D MRI brain-tumor segmentation. It won 1st place in the BRATS 2018 challenge. The network is based on asymmetric FCN combined with residual learning [5,89]. However, it has another branch at the encoder endpoint to reconstruct the original input image, similar to the auto-encoder architecture, as shown in Fig. 13. The motivation for the additional auto-encoding branch is to include regularization to the encoder part. The author also leveraged a group normalization (GN) rather than a batch normalization which is more suitable when the batch size is small [110]. The results of this network have dice similarity coefficients (DSCs) of more than 70 % and Hausdorff distances of less than 5.91 mm for BRATS brain dataset.

Table 2 organizes various deep learning methods reviewed in this paper based on their underlying network architectures. The dimensionality in this table means the dimensionality of the convolution kernel used in the network.

| Author | Categories | Specific | modalities | Object | Dimension |
|---|---|---|---|---|---|
| Kamnitsas K *et al.* [77] | Patch-based | - | MRI | Brain | 3D |
| Pereira S *et al.* [78] | Patch-based | - | MRI | Brain | 2D |
| Havaei M *et al.* [79] | Patch-based | - | MRI | Brain | 2D |
| Zhang W *et al.* [80] | Patch-based | - | MRI | Brain | 2D |
| Ibragimov [26] | Patch-based | Using Markov random fields | CT | Head and Neck | 3D |
| Qin J *et al.* [47] | Patch-based | Super-pixel-based patches | CT | Liver | 2D |
| Moeskops P *et al.* [82] | Patch-based | Multiple patches | MRI | Brain | 2D |
| Nie D *et al.* [83] | FCN | - | MRI | Brain | 2D |
| Brosch T *et al.* [84] | FCN | - | MRI | Brain | 3D |
| Roth HR *et al.* [85] | FCN | U-Net | CT | Organs | 3D |
| Chlebus G *et al.* [86] | FCN | V-Net | CT | Liver | 3D |
| Ronneberger O *et al.* [87] | FCN | Original U-Net | EM | Cell | 2D |
| Milletari F *et al.* [88] | FCN | Original V-Net | MRI | Prostate | 2D |

| | | | | Xenopus | |
| Çiçek O *et al.* [91] | FCN | U-Net | EM | kidney | 3D |
| | | | | embryos | |
| Wang C *et al.* [92] | FCN | U-Net | CT/MRI | Heart | 3D |
| | | | | Polyp, liver, | |
| Zhou Z *et al.* [93] | FCN | U-Net | CT/EM | lung, and | 2D/3D |
| | | | | cell | |
| Casamitjana A *et al.* [94] | FCN | V-Net | MRI | Brain | 3D |
| Dou Q *et al.* [96] | Cascaded | FCN-CNN | MRI | Brain | 3D |
| Christ *et al.* [97] | Cascaded | FCN-FCN | CT/MRI | Liver | 3D |
| Stollenga MF *et al.* [99] | Others | RNN | EM/MRI | Brain, cell | 3D |
| Yang X *et al.* [100] | Others | RNN | Ultrasound | Prostate | 2D |
| Chen J *et al.* [101] | Others | RNN/CNN | EM | Cell | 2D/3D |
| Men K *et al.* [107] | Others | Pyramid pooling | CT/MRI | Prostate | 2D |
| Mazdak AS *et al.* [108] | Others | Pyramid pooling | CT | Brain | 2D/3D |
| Myronenko [109] | Others | Auto encoder regularization | MRI | Brain | 3D |

**Table 2.** Categorized segmentation methods reviewed in this paper.

## 4.3. Implementation of deep learning

### 4.3.1. Framework & library

To develop and train deep learning networks, dedicated software frameworks and libraries have been developed in recent years. The number of such frameworks and libraries are growing rapidly and providing an exhaustive list is difficult. Consequently, we focus on the well-known open-source frameworks and libraries favored by deep learning practitioners:

• **Caffe**: is one of the early deep learning frameworks. It was developed by Berkeley Vision and Learning Center [111]. It is written in C++ with Python & MATLAB bindings for training and deploying.

• **Tensorflow**: is one of the most popular machine learning frameworks [112], developed by the Google Brain team based on Python language.

• **Torch**: is another deep learning framework to build complex network models. Early Torch was not based on Python language, but recently, Pytorch has been developed to extend to Python language [113].

• **Keras**: is one of open-source library for deep learning. It is written in Python language and it can be used on different frameworks such as Tensorflow, Microsoft Cognitive Toolkit, Theano, or PlaidML.

### 4.3.2. Segmentation Datasets

There are several datasets that are widely used for segmentation and are publicly available. For brain, brain tumor segmentation (BRATS), ischemic stroke lesion segmentation (ISLES), mild traumatic brain injury outcome prediction (mTOP), multiple sclerosis segmentation (MSSEG), neonatal brain segmentation (NeoBrainS12), and MR brain image segmentation (MRBrainS) dataset are available. The lung image database consortium image collection (LIDC-IDRI) consists of diagnostic and lung cancer screening thoracic CT scans with marked-up annotated lesions. For liver, there are public dataset of liver tumor segmentation (LiTS), 3D image reconstruction for comparison of algorithm database (3Dircadb), and segmentation of the liver (SLIVER07). Prostate MR image segmentation (PROMISE12) and automated segmentation of prostate structures (ASPS) dataset can be used for prostate segmentation. There is segmentation of knee image (SKI10) dataset for knee and cartilage as well. Brief explanations and categorization of each dataset are listed in Table 3. There may be more public dataset for segmentation not introduced in this review.

| Dataset | Modalities | Objects | URL |
|---|---|---|---|
| BRATS | MRI | Brain | https://www.med.upenn.edu/sbia/brats2018/registration.html |
| ISLES | CT/MRI | Brain | http://www.isles-challenge.org/ |
| mTOP | MRI | Brain | https://www.smir.ch/MTOP/Start2016 |
| MSSEG | MRI | Brain | https://portal.fli-iam.irisa.fr/msseg-challenge/data. |
| NeoBrainS 12 | MRI | Brain | http://neobrains12.isi.uu.nl/ |
| MRBrainS | MRI | Brain | http://mrbrains13.isi.uu.nl/ |
| LIDC-IDRI | CT | Lung | https://wiki.cancerimagingarchive.net/display/Public/LIDC-IDRI |
| LiTS | CT | Liver | https://competitions.codalab.org/competitions/17094 |
| SLIVER07 | CT | Liver | http://www.sliver07.org/ |
| 3Dircadb | CT | Body organs | https://www.ircad.fr/research/3dircadb/ |
| PROMISE12 | MRI | Prostate | https://promise12.grand-challenge.org/ |
| ASPS | MRI | Prostate | https://wiki.cancerimagingarchive.net/display/Public/NCI-ISBI+2013+Challenge+-+Automated+Segmentation+of+Prostate+Structures |
| SKI10 | MRI | Knee | http://www.ski10.org/ |

**Table 3.** Public dataset for segmentation.

## 5. Outlook and Discussion

5.1. Challenges and future research directions

Various deep learning networks offer great results for the medical image segmentation. In addition, the results from deep learning are comparable to those from manual segmentation by experts. in the case of HaN OAR segmentations, Nikolov *et al.* [46] showed that DSC values of brain segmentations were 95.1 % and 96.2 % from deep learning and manual, respectively. In the case of the cochlea, the segmentation accuracy of a deep learning methods is 97.8 % which is better than the 92.0 % accuracy obtained from the manual segmentation. Qin *et al.* [47] compared the liver segmentation results using the deep learning, active contouring, and the graph cut. He showed that the deep learning achieves 97.31 % accuracy compared to 96.29 % from active contouring, and 96.74 % from the graph cut.

Despite significant improvements achieved in segmentation of medical images using deep learning techniques, there are still some limitations pertaining to the issue of inadequate training datasets. In the public domain, it is often challenging to find accessible, high quality medical image data [67]. Without sufficient training samples, deep architectures with the expressiveness of ResNet [89], AlexNet [102], VGGNet [114], and GoogLeNet [115], often dramatically overfit the dataset, even with generic regularization strategies such as dropout [103], sparse regularization of the network [116], and model averaging [117]. Cho *et al.* [118] reported that the accuracy of CNN with GoogLeNet architecture for classification problems in medical image dataset was consistently improved after increasing training-dataset size. The classification task used in Cho's study is too simple to apply to realistic medical image processing such as segmentation; however, the study noted an important relation between performance and size of training dataset. The simplest way to increase the size of dataset is to transform the original dataset with random translation, flipping, rotation, and deformation. This concept, known as data augmentation, is already commonly used in classical machine learning algorithms. The effect of data augmentation is to mitigate the overfitting problem by enlarging the input dataset [119]. Deformation can be applied to data augmentation as well, introduced by Zhao *et al.* [120], and they successfully applied it to prostate radiation therapy [121].

Recent studies have used a deep learning concept of generative adversarial network (GAN) [122] to generate synthetic data from the training dataset [123-125]. In GAN, as shown in Fig. 14, two competing models (stages) are simultaneously trained. One stage is trained to generate data from noise input, and the other is trained to discriminate between synthesized data and real data. The generator in GAN tries to generate data that has a similar distribution to the original data, while the discriminator in a GAN tries to distinguish the two. Finally, competition of the two stages converges to where the discriminator cannot discriminate the original data from the synthesized data. The training process of a GAN involves training of the discriminator and generator sequentially. While the generator is fixed, the discriminator is trained on inputs from real dataset first and on inputs from the fixed generator later. The generator is then trained and updated under the fixed discriminator that is not updated during this time. Recently, to cope with requiring large amounts of manually annotated data for deep learning in segmentation, unsupervised deep learning models have received a great deal of attention, see, *e.g.*, [126].

Graph Neural Networks (GNNs) are useful tools on non-Euclidean domain structures (*e.g.*, images), which are being studied in recent researches [127]. Graphs are a kind of data structures that are composed of nodes and edges (or features and relationships). Graph-based expression have been received more and more attention due to their great expressive power for underlying relationships among data. Scarselli *et al.* [128] first introduced GNNs and directly applied existing neural networks to graph domain. There are several variants of GNNs with respect to graph types and propagation types. Zhou *et al.* [127] showed some applications including semantic segmentation in their review paper. GNNs can also be a useful tools for biomedical image segmentation because graph-structured data is more efficient where the boundaries are not grid-like and non-local information is needed.

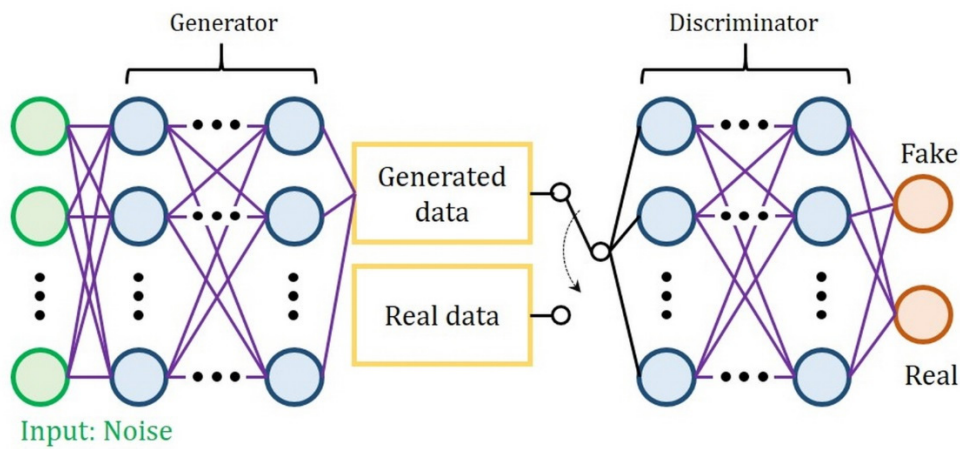

**Figure 14. Structure of the Generative Adversarial Network (GAN).**

Processing volumetric data via 3D convolutions using deep learning segmentation methods usually requires huge memory and long training time. In contrast, applying deep learning to 2D slice images often loses full 3D information. So, segmentation methods based on 2.5D that contains partial 3D volumetric information such as, an input data as several slice images, orthogonal images (transverse, sagittal, and coronal) at target location, maximum or minimum intensity projection (MIP or mIP) have been introduced [129-131].

Recent studies on medical image segmentation are primarily focused on the deep learning paradigm. Nevertheless, there are opportunities for further improvement of classical machine learning algorithms. For instance, in most classical machine learning algorithms, the feature extraction process is often carried out via a set of pre-specified filters. Therefore, devising data-driven feature extraction mechanisms for classical machine learning algorithms would significantly improve their performance as shown by Linsin *et al.* [132].

Current deep learning networks require a lot of hyperparameter tuning. Small changes in the hyperparameters can results in disproportionately large changes in the network output. Though the weights of the network are often determined automatically by back-propagation and stochastic gradient descent methods, many hyperparameters, such as the number of layers, regularization coefficients, and dropout rates, are still empirically chosen. Although relevant works have been studied to avoid problems that arise with these heuristic decisions [133,134], deep learning methods are not yet fully optimized. There are still many clinical problems to be solved. Moving forward, thoughtful consideration of the potential limitations of deep learning methodologies is extremely important.

**ACKNOWLEDGMENT**

This work was partially supported by NIH/NCI (1R01 CA176553), Varian Medical Systems, a gift fund from Huiyihuiying Medical Co, and a Faculty Research Award from Google Inc.